\documentclass[%
 reprint,
 amsmath,amssymb,
 aps,
]{revtex4-2}

\usepackage{graphicx}
\usepackage{dcolumn}
\usepackage{bm}
\usepackage{amsmath}

\def\be{\begin{equation}}
\def\ee{\end{equation}}
\def\bea{\begin{eqnarray}}
\def\eea{\end{eqnarray}}
\def\v{\vec}

\begin{document}

\preprint{APS/123-QED}

\title{Faraday Waves in a Low-Viscosity Fluid Covered with a Floating Elastic Sheet}

\author{Vahideh Sardari}
\email{v.sardari@iasbs.ac.ir}
\affiliation{Department of Physics, Institute for Advanced Studies in
Basic Sciences (IASBS), Zanjan 45137-66731, Iran}

\author{Leila Bahmani}
\email{l.bahmani86@gmail.com}
\affiliation{Department of Physics, Institute for Advanced Studies in
Basic Sciences (IASBS), Zanjan 45137-66731, Iran}

\author{Maniya Maleki}
\email{m\_maleki@iasbs.ac.ir}
\affiliation{Department of Physics, Institute for Advanced Studies in
Basic Sciences (IASBS), Zanjan 45137-66731, Iran}

\date{\today}

\begin{abstract}
The standing surface waves in a rectangular vertically oscillating vessel filled with water (Faraday waves) in the presence of a floating elastic sheet are studied experimentally and theoretically. The threshold amplitude of the instability and the wavelength of the patterns are measured as a function of the frequency. A theoretical model based on Hamiltonian method is used to describe the system. Using the experimental attenuation coefficients, we see a very good agreement between the theory and the experiment for threshold amplitude. Also, the dispersion relation obtained from the theory is consistent with experiments.

\end{abstract}

\maketitle


\section{Introduction}
When a vessel containing a liquid is subjected to vertical oscillation, nonlinear standing waves form on the liquid interface, which were first reported by Michael Faraday \cite{Faraday} and are called "Faraday Waves" after him. Many studies have been performed on Faraday waves, investigating their patterns \cite{patterns,patterns2,patterns3,patterns4,patterns5}, threshold amplitude for their formation, and their wavelength \cite{ampl1,patterns2, ampl2}. Other studies have investigated the effect of different parameters like the viscosity \cite{viscous,viscous2,viscous3} and surface tension \cite{tension,tension2} of the liquid, the filling depth, and the shape of the container \cite{viscous,depth,shape,shape2,shape3,shape4}.  
Several theories have been developed for the formation mechanism of the Faraday waves, namely the linear theory of an ideal fluid by Benjamin and Urcell \cite{Urcell}, the linear theory of a viscous liquid by Kumar and Tuckerman \cite{Kumar}, and the Lagrangian method of John Miles \cite{Miles67,Miles84,Miles90,Miles90A,Miles93}. Experimental studies have been also performed on the subject \cite{Miles90,viscous,1D,Douady, EN}. 

In this paper, we have studied Faraday waves in a rectangular vessel filled with water, to which we added a floating thin elastic sheet. This eliminates the effect of surface tension and introduces elastic effects instead. Waves forming on floating elastic sheets have been studied in different researches, and have applications in stability analysis of ocean ice sheets or floating constructions \cite{float0, float1, float2, float3, float4}. We have studied the standing waves on a floating elastic sheet, which covers almost completely the water's surface but is not pinned to the vessel. The presence of the sheet affects the dynamics of the system and changes the wavelength and threshold amplitude of the waves. We first propose a theoretical model to predict the wavelength and amplitude of the waves as a function of driving frequency. Then, we present the experimental results on the same problem and compare them with the theory. 

\begin{figure}[h!]
\centering
\centerline{\includegraphics[width=0.99\columnwidth]{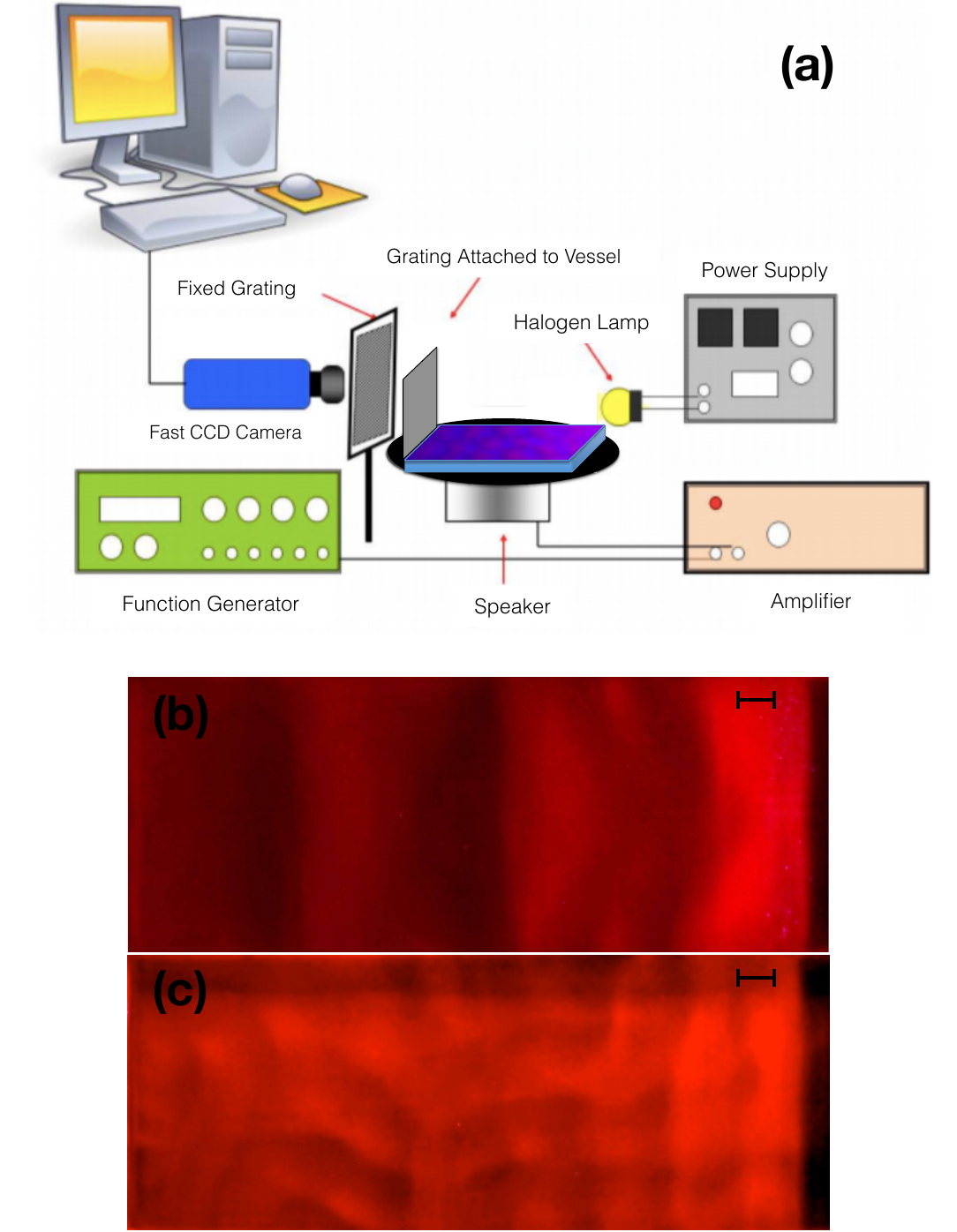}}
\caption{(a): Sketch of the experimental setup. (b) and (c): The Faraday wave patterns for two different frequencies: (a) 8 Hz and (b) 25 Hz . The scale bars are 2 cm long.}
\label{Schematic}
\end{figure}

\section{Theory}
\subsection{Lagrangian}
For modeling the system, we use the Lagrangian method used by John Miles \cite{Miles90, Miles90A}.  Considering a vorticity-free incompressible flow, the velocity field $\vec{v}$ with respect to the oscillating liquid container can be written as the gradient of a scalar field $\phi(x,y,z;t)$, where $x$, $y$ and $z$ are the Cartesian coordinates and $t$ is time. $\phi$ is defined in the volume of the liquid, i. e.
\begin{equation}
\vec{r}=(x,y)\in S,\\
-d<z<\eta,
\end{equation}
where $S$ is the container cross section (a rectangle in our case), and $\eta(\vec{r};t)$ gives the vertical displacement $z$ of the free surface points with respect to the plane of the level surface. $\phi$ should satisfy
\begin{eqnarray}
\vec{v}=\vec{\nabla}\phi,\\
\nabla^2\phi=0.
\end{eqnarray}

For standing waves, we can separate the local and temporal variables of $\eta$ and $\phi$ as follows
\begin{eqnarray}
\eta(\vec{r};t)&=&\sum_n \eta_n(t) \psi_n(\vec{r}),\\
\phi(\vec{r},z;t)&=&\sum_n \phi_n(t) \chi_n(\vec{r},z),
\end{eqnarray}
where $\eta_n$ and $\phi_n$ are generalized coordinates, and $\psi_n$ and $\chi_n$ are orthogonal eigenfunctions in the linear approximation \cite{Miles90}.

The normal velocity on the bottom and walls of the container must be zero. Thus, we have the boundary condition
\begin{equation}
\hat{n} . \vec{\nabla} \phi=0
\end{equation}
on this area. The dynamic boundary condition on the free surface reads

\begin{equation}
v_z=\frac{\partial \phi}{\partial z} = \frac{\partial \eta}{\partial t} + \vec{v} . \vec{\nabla} \eta.
\end{equation}

In the linear approximation, the last term is negligible and we can write the equation to the leading term as
\begin{equation}
\frac{\partial \phi}{\partial z} \approx \frac{\partial \eta}{\partial t} ,\; \; \;   \left( z=\eta(\vec{r})\approx 0\right).
\end{equation}

Considering the container to have a rectangular cross-section with walls at $x=0, x=L_x$; $y=0, y=L_y$ and $z=-d$, one can write $\phi$ as a Fourier series and obtain the from of functions $\chi_n$ as
\bea
\begin{split}
\chi_n &= A_n \cos(k_{nx} x) \cos(k_{ny} y) \times\\
          &\text{ sech} (k_n d) \cosh k_n (z+d),\\
          \v{k}_n &= (k_{nx},k_{ny})=(\frac{n\pi}{L_x},\frac{n\pi}{L_y}). \label{chi}
\end{split} 
\eea

 Substituting equation 9 in equation 8, we will have

\begin{equation}
(\nabla^2+k_n^2)\psi_n=0,   
\end{equation}

Using the variational principle, one can write
\begin{equation}
\phi_m=\sum_{j,l} \kappa^{-1}_{ml}d_{jl} \dot{\eta}_j,   
\end{equation}
where
\bea
d_{mn}&=&S^{-1} \iint dS (\chi_n)_{z=\eta}\psi_m,\\
\kappa_{mn}&=&S^{-1} \iint dS \int dz \vec{\nabla} \chi_m.\vec{\nabla} \chi_n.
\eea
Then we can calculate the Lagrangian of the system as a function of generalized coordinates $\eta_n$ by writing $T_0$, the kinetic energy and $V_0$, the potential energy as  \cite{Miles67} 
\bea
T_0&=&\frac{1}{2} \sum_{m,n}\rho S a_{mn} \dot{\eta}_m \dot{\eta}_n,\\
V_0&=&\frac{1}{2}\sum_n \rho S  g' \eta_n^2,
\eea
where $\rho$ is the fluid density, $g'$ is the sum of gravitational acceleration $g$ and container acceleration $\ddot{z}_0$, and $\pmb{a}$ is a symmetric matrix defined as 
\begin{equation}
\pmb{a}=\pmb{d} \pmb{\kappa^{-1}}\pmb{d'},
\end{equation}
 The elements of $\pmb{a}$ can be expanded as taylor series of $\eta_n$.
 \begin{equation}
 a_{mn}=\delta_{mn} a_m+\sum_{l}a_{lmn}\eta_l+\frac{1}{2} \sum_{j,l} a_{jlmn}\eta_j \eta_l+... ,
 \end{equation}

where 
\begin{equation} \label{an}
a_n=k_n^{-1} \coth k_nd \equiv \frac{g}{\omega_n^2},   
\end{equation}
and other series coefficients are related to correlation integrals $C$ and $D$ 
\bea
a_{lmn}&=&C_{lmn}-D_{lmn}a_m a_n,\\
a_{jlmn}&=&-D_{jjlmn}(a_m+a_n) +2 D_{jmi} D_{lni} a_l a_m a_n,
\eea
which are defined as follow \cite{Miles90}.
\bea
C_{lmn}&=&S^{-1} \iint dS \psi_l \psi_m \psi_n dS, \\
C_{jlmn}&=&S^{-1} \iint dS \psi_j \psi_l \psi_m \psi_n dS,\\
D_{lmn}&=&S^{-1} \iint dS \psi_l \vec{\nabla}\psi_m. \vec{\nabla}\psi_n dS, \\
D_{jlmn}&=&S^{-1} \iint dS \psi_j \psi_l \vec{\nabla}\psi_m. \vec{\nabla}\psi_n dS.
\eea
Based on these equation, the Lagrangian is calculated as
\begin{multline}
\mathcal{L_0}=\frac{1}{2}\sum_n a_n(\dot{\eta}_n^2-\omega_n^2 \eta_n^2 - \ddot{z}_0\eta_n^2)+\frac{1}{2} \sum_{l,m,n} a_{lmn}\eta_l \dot{\eta}_m \dot{\eta}_n\\
+\frac{1}{4} \sum_{j,l,m,n} a_{jlmn} \eta_j \eta_l \dot{\eta}_m \dot{\eta}_n.
\end{multline}

Now, we should incorporate the terms related to the elastic floating sheet sheet into the Lagrangian. The elastic potential of the sheet reads
\begin{equation}
\begin{split}
V_{el}&=\frac{1}{2} \iint K [\nabla^2\eta]^2 dS\\
         &=\frac{1}{2} \sum_{m,n} \iint  K [\nabla^2\psi_m] [\nabla^2\psi_n] \eta_m(t) \eta_n(t) dS\\
         &=\frac{1}{2} \sum_{n}K k_n^4 \eta_n^2 S,
\end{split}
\end{equation}
where $K$ is the bending rigidity of the sheet.

The potential energy due to the vertical displacement of the sheet can be written as
\begin{equation}
\begin{split}
V_{g}&= \frac{1}{2} \iint \sigma [1+ \lvert \vec{\nabla} \eta \rvert ^2]^{1/2} \eta g' dS \\
           &=\frac{\sigma}{2} \sum_{l,m,n} \iint  g'   \v{\nabla}\psi_m . \v{\nabla}\psi_n  \eta_m(t) \eta_n(t) \psi_l(t) \eta_l(t) S\\
           &=\frac{\sigma}{2} \sum_{l,m,n} g' D_{lmn} \eta_l (t) \eta_m (t) \eta_n (t),
\end{split}
\end{equation}
where $\sigma$ is the mass per unit area of the sheet.

The Kinetic energy of the sheet is given by
\begin{equation}
\begin{split}
T_{s}&= \frac{\sigma}{2} \iint  [1+ \lvert \vec{\nabla} \eta \rvert ^2]^{1/2} (\v{\nabla}\phi)^2_{z=\eta} dS \\
          &=\frac{\sigma}{2} \sum_{m,n} \iint \v{\nabla}\chi_m . \v{\nabla}\chi_n dS + \frac{\sigma}{4}\sum_{j,l,m,n} \iint  \v{\nabla}\chi_m . \v{\nabla}\chi_n \times \\
          & \v{\nabla}\psi_j . \v{\nabla}\psi_l \eta_j(t) \eta_l(t)  \phi_m(t) \phi_n(t)  dS\\
          &=\frac{\sigma}{2} \sum_{m,n}S \phi_m \phi_n \kappa'_{mn}+ \frac{\sigma}{4} \sum_{j,l,m,n} S M_{jlmn}\phi_m \phi_n \eta_j  \eta_l ,
\end{split}
\end{equation}
where 
\begin{equation}
M_{jlmn}=k_n^2 k_l^2 \eta_j \eta_l C_{jlmn} + D_{mnjl} \kappa_m \kappa_n.
\end{equation}
The elements of $\pmb{\kappa'}$ are defined as 
\begin{equation}
\kappa'_{mn}=\frac{1}{S} \iint \v{\nabla}\chi_m . \v{\nabla}\chi_n dS,
\end{equation}
and be expanded in power series of $\eta$ as follows
\bea
\kappa'_{mn}&=&\delta_{mn} (k_m^2+\kappa_m^2) + \sum_{l}[D_{lmn}(\kappa_m+\kappa_n)   \nonumber \\
&+&  C_{lmn} (k_m^2 \kappa_n+ k_n^2\kappa_m)]\eta_l+\frac{1}{2}\sum_{j,l} [D_{jlmn}(k_m+\kappa_n T_m ) \nonumber\\
&+ &C_{jlmn} k_n (k_m^2+\kappa_m \kappa_n)]\eta_j \eta_l + ... ,\\
\kappa_m &=&a_m^{-1}=k_m \tanh k_m d \equiv k_m T_m.
\eea
 Using equation 11, $T_s$ can be written as
 \begin{equation}
T_s=\frac{\sigma}{2}\sum_{m,n} b_{mn}\dot{\eta}_m \dot{\eta}_n +\frac{\sigma}{4}\sum_{j,l,m,n}r_{jlmn} \eta_j \eta_l \dot{\eta}_m \dot{\eta}_n  .
\end{equation}
The matrix $\pmb{b}$ and the coefficient $r$ are defined as follows
\bea
b_{mn}&=&\sum_{j,l}\kappa'_{mj} \kappa^{-1}_{jl} a_{ln},\\
r_{jlmn}&=& \kappa^{-1}_{mn}M_{jlmn}a_{mn}.
\eea
 The elements of $\pmb{b}$ can also be expanded as
 \bea
b_{mn}&=&(k_n^2a_n^2+1) \delta_{mn} + \sum_{l} b_{lmn} \eta_l \nonumber \\
&+& \sum_{j,l} b_{jlmn} \eta_j \eta_l,\\
b_{lmn}&=&(k_n^2a_n^2+1)(a_{lmn} \kappa_n-1) \nonumber \\
&+&2 (D_{lmn}+C_{lmn}k_n^2),\\
b_{jlmn}&=&\frac{1}{2} a_{jlmn} (\kappa_n+k_n^2 a_n) + D_{jlmn} a_n (k_n a_n +T_n)\nonumber \\
&+& C_{jlmn}k_n (k_m^2a_n^2+1).
\eea
Since in equation 33 the last term has is $r_{jlmn}\times O(\eta^4)$,we keep only the leading term of $r_{jlmn}$, which is

\begin{equation}
r_{jlmn}= 2 (D_{jlmn}+C_{mlj} k_n^2 k_j^2 a_n^2) \delta_{mn}.
\end{equation}

Defining $\gamma \equiv \sigma/\rho$, the Lagrangian of the system with the elastic sheet divided by $\rho S$ is hence given by
\bea
\mathcal{L}&=&\frac{1}{\rho S}(T-V) \nonumber \\
&=&\frac{1}{2}\sum_{j,l,m,n}\left (a_{mn}+\gamma b_{mn} 
+\frac{\gamma}{2} r_{jlmn} \eta_j \eta_l\right) \dot{\eta}_j \dot{\eta}_l \nonumber\\
& -& \frac{1}{2} \sum_{m,n} \left( (g' \delta_{mn} + \frac{K k_n^4}{\rho}\right)\eta_m \eta_n.
\eea
Substituting the expansions and keeping up to the second order, we have
\begin{equation}
\begin{split}
\mathcal{L}&=\frac{1}{2}\sum_n   \frac{a_n}{2} \left[ \left( 1+\gamma k_n^2 a_n \right) \dot{\eta}_n^2 - \omega_n^2\eta_n^2\right]  \\
&- \frac{1}{2} \sum_n  \left[ \left( \ddot{z}_0+\frac{Kk_n^4}{\rho}\right) \eta_n^2+ \gamma \dot{\eta}_n^2 \right] 
\end{split}
\end{equation}

\subsection{Threshold Amplitude}
Miles showed that if the reservoir oscillates vertically with the angular velocity $2 \omega$ and amplitude $a_0$
\begin{equation}
z_0=a_0 \cos 2\omega t,
\end{equation}
for low amplitudes $\omega^2 |a_0|\ll g$ in the nonlinear regime, only the harmonic and sub-harmonic responses are excited, with slowly varying amplitudes \cite{Miles93}. Hence, each mode $\eta_n$ can be written as
\begin{equation}
\begin{split}
\eta_n &=\delta_{1n} \lambda \left[ p(\tau) \cos \omega t + q(\tau) \sin \omega t \right] \\
           &+\frac{\lambda^2}{a_1} \left[A_n(\tau) \cos 2\omega t + B_n(\tau)  \sin 2\omega t +C_n(\tau) \right],
\end{split}
\end{equation}
where $p$, $q$, $A_n$, $B_n$ and $C_n$ are the slowly varying amplitudes and
\bea
\epsilon = \frac{a_0}{a_1}, \quad 0<\epsilon \leqslant 1,\\ \label{epsilon}
\tau=\epsilon \omega t, 
\eea
and $\lambda=O(\epsilon^{1/2} a_1)$ is a length scale. Putting this in the Lagrangian, we can write the action integral in one period and divide it to the period to find
\bea
\langle \mathcal{L}\rangle &=& \frac{1}{2}\epsilon g \lambda ^{2} \bigg[ \beta(p^2+q^2) + \left ( 1+\gamma(a_1k_1^2+a_{1}^{-1}) \right ) \nonumber \\
&\times & (\dot{p}q-\dot{q}p) + \left(p^2-q^2\right)  \bigg] \nonumber \\
&+& \frac{g \lambda^4}{2 a_1^2} \bigg[ \frac{1}{16} a_1 (p^2+q^2)^2 ( a_{1111} + \gamma b_{1111} + \frac{\gamma}{4} r_{1111} )  \nonumber\\
 &+& \sum_n \bigg\{ \frac{\Omega_n}{2}(A_n^2+B_n^2) -C_n^2 (1+k_n^4 l_*^2) \\
 &+&\left( a_{11n} + \frac{\gamma}{2}(b_{11n}+b_{1n1}) - \frac{1}{4} (a_{n11}+\gamma b_{n11})\right) \nonumber\\
 &\times& \left( A_n (p^2-q^2)+2B_n p q \right) +\frac{C_n}{2}(a_{n11}+\gamma b_{n11})(p^2+q^2) \bigg]. \nonumber
\eea

In the above equation, dot means the derivative with respect to $\tau$ and parameters $\beta$, $l_*$ and $\Omega_n$ are defined as follows.
\bea
\beta&=&\frac{1}{2\epsilon} \left[ \frac{ \left(1+\gamma(a_1k_1^2+a_1^{-1})\right) \omega^2 - \omega_1 ^2}{\omega_1 ^2} -k_1^4l_*^4 \right], \\ \label{beta}
\Omega_n&=&  4 \frac{a_n}{a_1} \left(1+\gamma (a_n k_n^2+a_n^{-1}) \right) - k_n^4 l_*^4 -1 ,\\
l_*&=& \left( \frac{K}{\rho g} \right)^{1/2}, \label{l*}
\eea

Using the principle of stationary action, $\langle \mathcal{L}\rangle$ should not change with variation of amplitudes $p$, $q$, $A_n$, $B_n$
and $C_n$ \cite{Miles93}. This condition gives the amplitudes as
\bea
A_n&=&\frac{-(p^2-q^2) \left( a_{11n}+\frac{\gamma}{2}(b_{11n}+b_{1n1}) -\frac{1}{4}(a_{n11}+\gamma b_{n11})\right)}{\Omega_n}, \nonumber\\
B_n&=&\frac{-2pq \left( a_{11n}+\frac{\gamma}{2}(b_{11n}+b_{1n1}) -\frac{1}{4}(a_{n11}+\gamma b_{n11})\right)}{\Omega_n}, \nonumber\\
C_n&=&\frac{(a_{n11}+\gamma b_{n11})(p^2+q^2)}{4(1+k_n^4l_*^2)}.
\eea
Substitution of the amplitudes in the average Lagrangian, we obtain
\bea
\langle \mathcal{L}\rangle&=&\epsilon g \ell^2 \bigg\{\frac{1}{2}(\dot{p} q - p \dot{q}) \left[ 1 + \gamma (a_1 k_1^2 + a_1^{-1} \right] + H(p,q) \bigg\}\nonumber\\
&+& O(\epsilon^2),
\eea
where the length scale $\ell$ and the Hamiltonian $H$ are defined as follows
\bea
&\ell&=2 \left(  \frac{\epsilon}{|f|} \right)^{1/2} k_1^{-1} \tanh k_1 d,\\
&H&(p,q)=\frac{1}{2} (\beta+1)p^2+ \frac{1}{2}(\beta-1)q^2+\frac{s}{4}(p^2+q^2)^2,
\eea
and $s$ is
\bea
s&=&\text{sgn} (f),\\
f&=&\frac{a_1 \tanh^4 k_1 d}{2} \Bigg\{ a_{1111}+\gamma b_{1111}+\frac{\gamma}{4} r_{1111} \nonumber  \\
&+& \frac{(a_{n11}+\gamma b_{n11})^2}{1+k_n^4l_*^2}\\
&+& \frac{\left[ 4a_{11n}+8\gamma (b_{11n}+b_{1n1})-(a_{n11}+\gamma b_{n11}) \right]^2}{8a_n \left[ a_1(1+k_n^4l_8^2)-1-\gamma (a_1k_1^2+a_1^1)\right]} \Bigg\}. \nonumber 
\eea
In order to take into account the viscous dissipation in the system, we have to add a dissipation function $D$ to the Hamiltonian
\begin{equation}
D=\frac{1}{2} \alpha (p^2+q^2), \quad \alpha=\frac{\delta}{\epsilon},
\end{equation}
in which $\delta$ is the damping ratio of a free wave of frequency $\omega$ \cite{Miles93}. The evolution equations are thus obtained as
\bea
&\dot{p}&+\alpha p + (\beta - 1 +p^2+ q^2)q=0,\\
&\dot{q}&+\alpha q - (\beta + 1 +p^2+ q^2)p=0.
\eea
For $\alpha>0$, the only fixed point of the above equations is the point $p=q=0$, which corresponds to the nonlinear response of the system as a sub-harmonic internal resonance  \cite{Miles93}. To determine whether the fixed point is stable or not, we apply a small disturbance 
$\propto \exp{\sigma \tau}$ and find the stability condition from the characteristic equation
\begin{equation}
\Delta=
\begin{vmatrix}
      \sigma+\alpha+2pq & \beta-1+p^2+3q^2\\
-(\beta+1+q^2+3p^2)&\sigma + \alpha - 2pq \end{vmatrix} =0.
\end{equation}
For $p=q=0$, the solution of the characteristic equation is
\begin{equation}
\sigma=-\alpha+(1-\beta^2)^{1/2},
\end{equation}
and $\sigma=0$ gives the neutral stability condition
\begin{equation}
\beta=\pm (1-\alpha^2)^{1/2}.
\end{equation}
This condition gives us the threshold amplitude $a_0$ for the Faraday waves. substituting equation \ref{an}, \ref{epsilon}, \ref{beta} and \ref{l*},  we find
\begin{widetext}
\begin{equation}
 a_0=(k \text{ tanh }k d)^{-1}\left\{\delta^2 + \left[ \frac{(1+2\gamma k \text{ coth }2k d)\omega^2 - \omega'^{2}}{2 \omega^2} - \frac{k^4 K}{2 \rho g} \right]^2 \right\}^{1/2},
 \label{a0}
\end{equation}
\end{widetext}
where 
\begin{equation}
\omega'=\omega (1- \delta).
 \end{equation}
This equation becomes equivalent to the threshold for Faraday waves without the sheet in the limit of $\sigma \rightarrow 0$ and $K \rightarrow 0$  \cite{Miles90} 
\begin{equation}
a_0=(k \text{ tanh }k d)^{-1}\left\{\delta^2 + \left[ \frac{\omega^2 - \omega'^2}{2 \omega^2}  \right]^2 \right\}^{1/2}
\end{equation}
As we expect, the threshold is higher in the presence of the elastic sheet, due to the effects of the weight and the elasticity of the sheet. 

\subsection{Dispersion Relation}

To find the dispersion relation of the waves, we first write the kinetic and potential energies of the system. Using equations 14,15, 26, 27 and 28  we can write

\bea
T&=&\frac{1}{2}\int \rho v^2 dV+\frac{1}{2}\int \sigma v^2 dA,\\
V&=& \frac{1}{2}\int \left[\rho g \eta^2 + K(\nabla^2 \eta)^2+2 \sigma g \eta \left(1+|\vec{\nabla} \eta|^2\right)^{1/2}\right] dS,\nonumber
\eea
where $dV$ is the volume element inside the liquid and $dA$ is the surface element on the sheet.
Using equation 2, we have
\begin{equation}
T=\frac {\rho}{2} \int  |\vec{\nabla} \phi|^2 dV+\frac {\sigma}{2} \int  \left[ 1+ (\vec{\nabla} \eta)^2 \right]^{1/2} |\vec{\nabla}\phi|_{z=\eta}^2 dS.
\end{equation}
Then, we use the Fourier transforms of the functions $\eta(\vec{r};t)$ and $\phi(\vec{r},z;t)$, which are the continuous forms of equations 4 and 5.

\begin{equation}
\eta(\vec{r};t)=\int \frac{d\vec{k}}{(2\pi)^2} \;\tilde{\eta}(\vec{k};t) e^{i\vec{k}.\vec{r}},
\end{equation}
and using 9 we have
\begin{equation}
\phi(\vec{r},z;t)=\int  \frac{d\vec{k}}{(2\pi)^2} \;\tilde{\phi}(\vec{k};t) \cosh k(z+d) \text{ sech }kd \; e^{i\vec{k}.\vec{r}}.
\end{equation}
Substituting the above transforms into the boundary condition 8, we obtain a relation between the Fourier amplitudes as
\begin{equation}
\tilde{\phi}(\vec{r};t)=\frac{\tilde{\eta}_t (\vec{r};t)}{k \text{ tanh }kd},
\end{equation}
where $\tilde{\eta}_t$ is the partial derivative of  $\tilde{\eta}$ with respect to $t$.
Writing the kinetic energy $T$ in terms of $\tilde{\eta}$ up to the second order in $\eta$, we derive
\bea
T&=&\frac{\rho}{2} \int  \frac{d\vec{k}}{(2\pi)^2} \frac{|\tilde{\eta}_t (\vec{k};t)|^2}{k \text{ tanh }kd}\\ \nonumber
  &+&\frac{\sigma}{2} \int  \frac{d\vec{k}}{(2\pi)^2}|\tilde{\eta}_t (\vec{k};t)|^2 (1+ \text{coth}^2 \;kd).
\eea
Likewise, the potential energy $V$ is calculated up to the second order as
\begin{equation}
V=\frac{1}{2} \int \frac{d\vec{k}}{(2\pi)^2} (\rho g+K k^4)  |\tilde{\eta} (\vec{k};t)|^2
\end{equation}
Now, we form the Lagrangian with the constraint that the total surface area of the sheet is constant, using a Lagrange multiplier $\Lambda$. 
\begin{equation}
L=T-V-\Lambda \int dA, 
\end{equation}
where
\begin{equation}
\int dA \simeq \int  \left( 1+\frac{|\vec{\nabla}\eta|^2}{2} \right) dS=\int \frac{d\vec{k}}{(2\pi)^2} \frac{k^2}{2}|\tilde{\eta} (\vec{k};t)|^2
\end{equation}
Thus the Lagrangian can be written as a Fourier transform as follows
\begin{widetext}
\bea
L&=&\int \frac{d\vec{k}}{(2\pi)^2} \left\{ \left[ \frac{\rho}{2k \text{ tanh } kd}+\frac{\sigma (1+ \text{coth }kd)}{2}\right]|\tilde{\eta}_t (\vec{k};t)|^2 - \frac{\rho g +Kk^4+\Lambda k^2}{2} |\tilde{\eta} (\vec{k};t)|^2 \right\} \\  \nonumber
   &\equiv &\int \frac{d\vec{k}}{(2\pi)^2} \left( l_1(k) |\tilde{\eta}_t (\vec{k};t)|^2 + l_2(k) |\tilde{\eta} (\vec{k};t)|^2 \right).
 \label{eq:wideeq}
\eea
\end{widetext}

The Euler-Lagrange equation for each Fourier mode then is written as
\begin{equation}
\frac{\partial L}{\partial \tilde{\eta} (\vec{k};t)}- \frac{d}{dt}\frac{\partial L}{\partial \tilde{\eta}_t (\vec{k};t)}=0,
\end{equation}
which gives
\begin{equation}
\tilde{\eta}_{tt}(\vec{k},t)=\frac{l_2(k)}{l_1(k)} \tilde{\eta}(\vec{k},t).
\end{equation}
Considering $\tilde{\eta}(\vec{k},t)$ to be a product of two functions of $t$ and $k$ 
\begin{equation}
\tilde{\eta}(\vec{k},t)=Q(t) R(k),
\end{equation}
and supposing $Q(t)$ to be a periodic function
\begin{equation}
Q(t) = e^{i \omega t},
\end{equation}
we obtain 
\begin{equation}
\omega^2=\frac{-l_2(q)}{l_1(q)}.
\end{equation}
Thus the dispersion relation reads
\begin{equation}
\omega = \sqrt{\frac{\rho g k + \Lambda k^3 +K k^5}{\sigma k +(\sigma k+\rho)\text{coth }kd }}. \label{disper}
\end{equation}
The above equation gives the relation for the non-dissipative case. In order to take into account the viscous dissipation, we should replace $\omega$ with $\omega'$ \cite{Miles90}.

We should also note that Eq. (\ref{disper})  is written in the continuous form \cite{Miles90}, although we have quantized values for $(k_x,k_y)$ as seen in Eq. (\ref{chi}) for a bounded rectangle reservoir. To take into account the quantization, $k$ should be replaced with allowed values of $k_n$ from Eq. 5.

\section{Experiments}

\begin{figure}[h!]
\centering
\centerline{\includegraphics[width=0.99 \columnwidth]{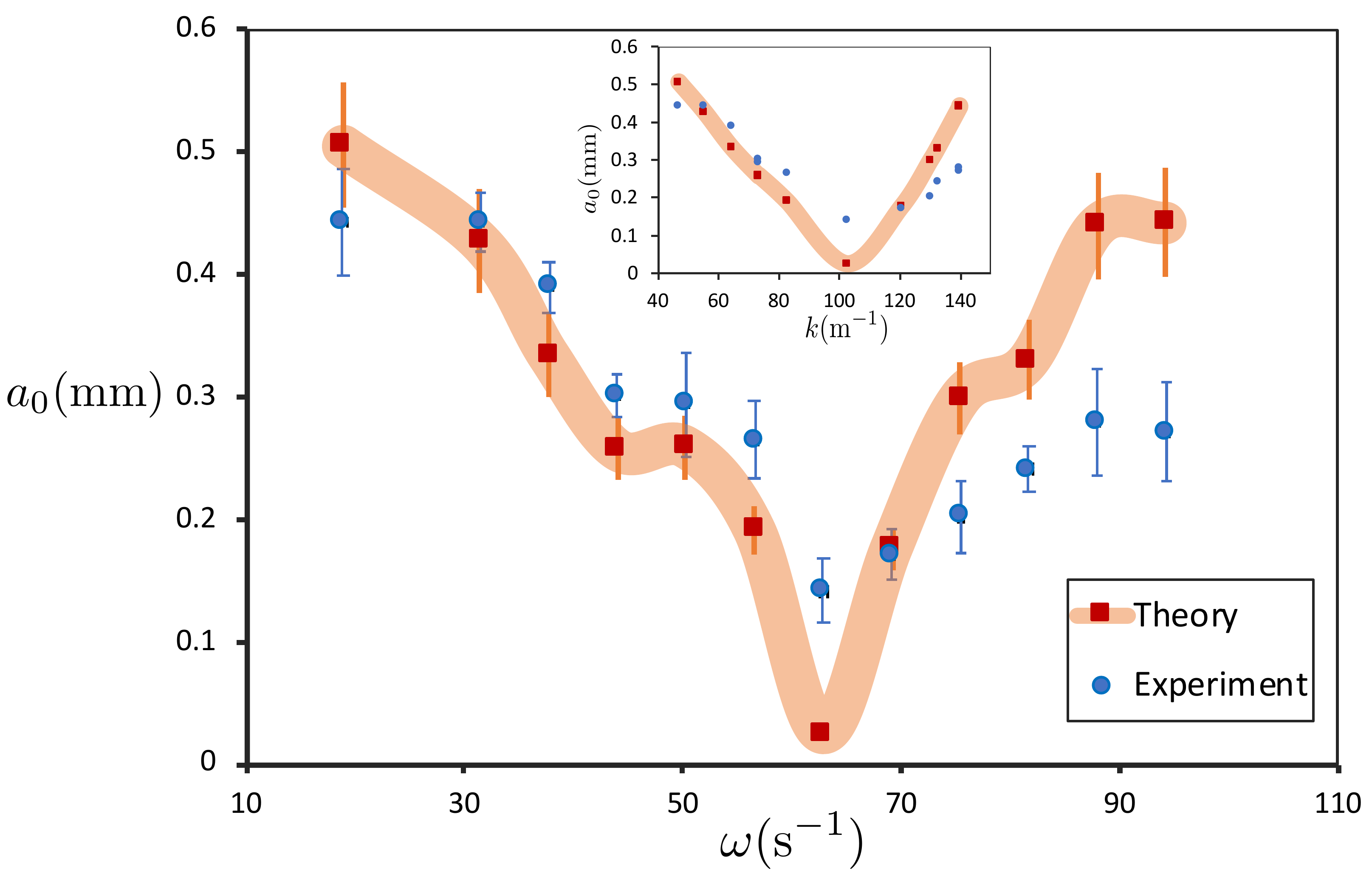}}
\caption{The threshold amplitude of the waves vs. the angular frequency. Blue circles: experimental data, and red squares: theoretical relation. Experimental error bars are calculated as the standard errors of different experiments and theoretical error bars are mainly due to the error in measuring the dissipation rate. Inset: the threshold amplitude vs. wavenumber. }
\label{amp}
\end{figure}

 We performed experiments to validate our theoretical predictions of threshold amplitude and wavelength. We used a rectangular vessel with dimensions of 33.5 cm $\times$ 14.1 cm and a depth of  1.5 cm. The vessel was filled with water and a floating rectangular silicone sheet with the same dimensions and a thickness of 0.25$\pm$0.02 mm was put on the water's surface. The vessel was placed on top of a horizontal speaker. We made oscillations with different frequencies and amplitudes using a function generator and an amplifier. The oscillation amplitude was measured accurately with the moir\'e technique (See Appendix) with an accuracy of 50 microns. 
To measure the threshold amplitude, we first fixed the oscillation frequency and then increased the amplitude gradually, until we observed the Faraday waves. We used a stroboscope to measure the frequency of the waves in each experiment. We changed the stroboscope frequency until we saw a fixed pattern on the surface. We observed that the frequency of the Faraday waves was always half the oscillation frequency. The experimental values of the threshold amplitude as a function of the driving frequency are shown in Fig. \ref{amp}. 

To compare the experimental results with the theory, we must know the experimental value for the bending rigidity $K$. We used the method used in \cite{Abou} and the measured value was $5.2\pm 0.5 \times 10^{-6}$ J (see Appendix for details). Other values used were the mass per unit area of the sheet $\sigma=$ $\text{0.284 Kg/m}^2$, the surface tension of water $\Lambda=0.073$ N/m and the density of water $\rho=1000 \text{ Kg/m}^3$. 

Another parameter we need to evaluate is the damping ratio $\delta$, defined as $\delta=\Gamma/\omega$, $\Gamma$ being the damping rate of the waves \cite{Miles84}. Henderson and Miles \cite{Miles90} showed that all the theoretical works give an unrealistic value for the damping rate; thus it is better to use the experimental value in the theoretical threshold formula. We used an almost vertical laser beam reflected from the surface, and projected the reflection on a vertical wall using a mirror. As shown in Appendix, Fig. 3 (a, b), when there is no Faraday wave formed, a light spot is seen on the wall, while a Lissajous-like curve forms in the presence of the waves. This enabled us to determine the exact threshold.  In order to determine the damping rate, we filmed the Lissajous curve with a high enough frame rate (300 or 600 fr/s) to locate the laser spot in different frames. After the Faraday waves were formed, we turned off the speaker and let the waves damp. As described in Appendix, the damping rate $\Gamma$ was then measured for each frequency and $\delta$ in Eq. (\ref{a0}) was obtained. 

\begin{figure}[ht!]
\centering
\centerline{\includegraphics[width=0.99 \columnwidth]{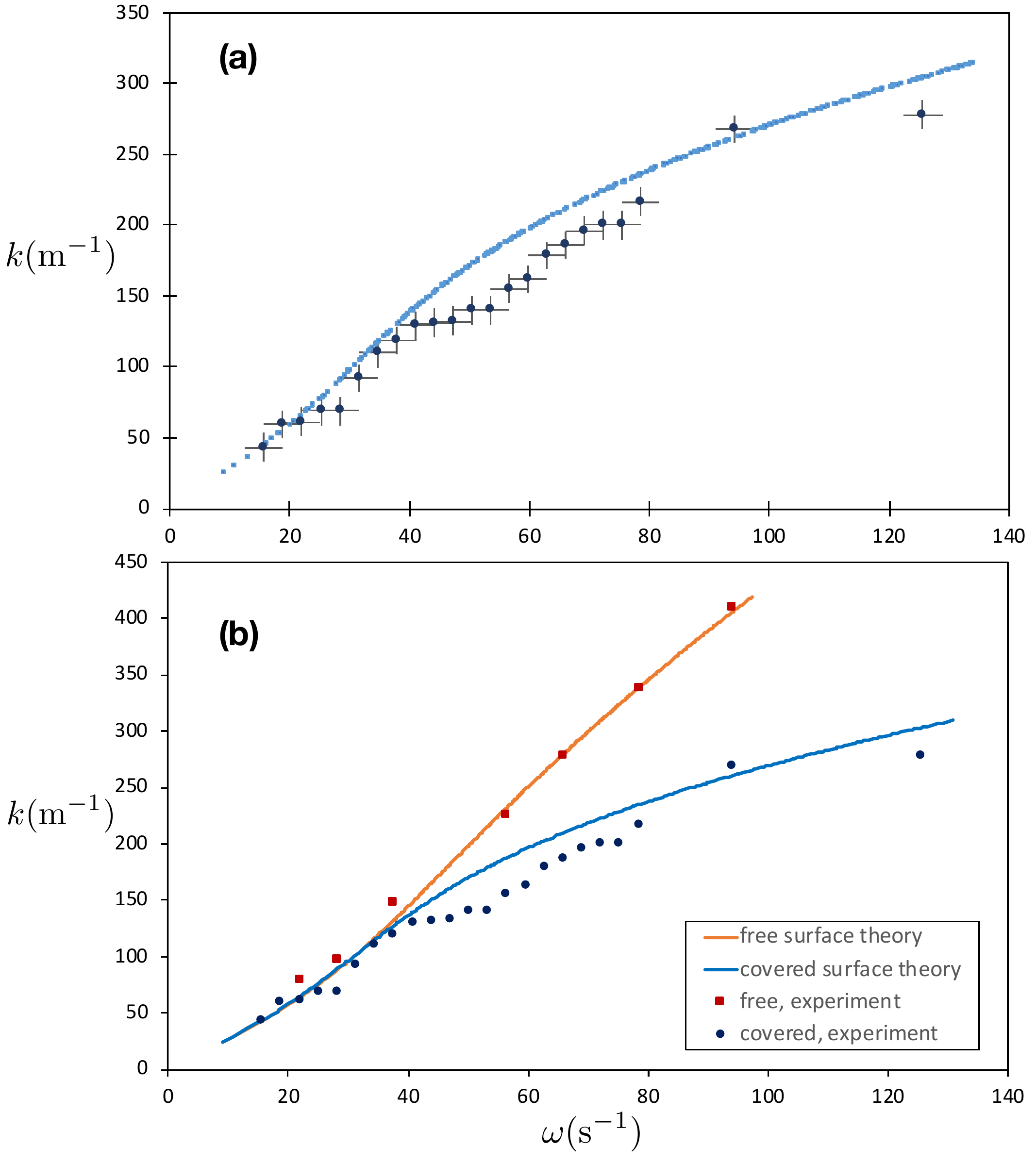}}
\caption{The wavenumber of the Faraday patterns as a function of the angular frequency. (a) Red circles: experimental data, and blue squares: quantized theoretical values. The error bars come from the uncertainties in determining $\omega$, $n_x$ and $n_y$. (b) Comparison of the free surface Faraday waves (without the elastic sheet) and covered surface waves. The symbols correspond the the experimental data and the curves represent the theoretical values.}
\label{dis}
\end{figure}

For measuring the wave number $k$, we used the method of Douady and Fauve \cite{DF}. We used the stroboscope as the light source to get a sharp picture and took pictures of the waves from above (Fig. 1). For low frequencies, the patterns were regular, and the wave numbers could be found easily (Figure 1(b)). For high frequencies, the patterns became irregular and the wavenumber was measured using the Fourier analysis (Fig. 1(c)). Having all the necessary parameters, we can calculate theoretical amplitudes for each frequency. The results are shown in Fig. \ref{amp}. There is a very good agreement between the theory and the experiment, except for forcing frequencies more than 18 Hz. In this range, the formation of capillary waves at the interface menisci between the water and the vessel, and between water and the sheet leads to wave formation in very small amplitudes \cite{capillary}. The minimum present in the theoretical and experimental data is similar to the behavior reported in \cite{Miles90A}, which is equivalent to the inset curve in Fig. 2. (Since we used experimental values for dissipation rate and put it in the theoretical formula, the theoretical values are not continuous and we do not have a smooth theoretical curve.).

\begin{figure}[h!]
\centerline{\includegraphics[width=0.7\columnwidth]{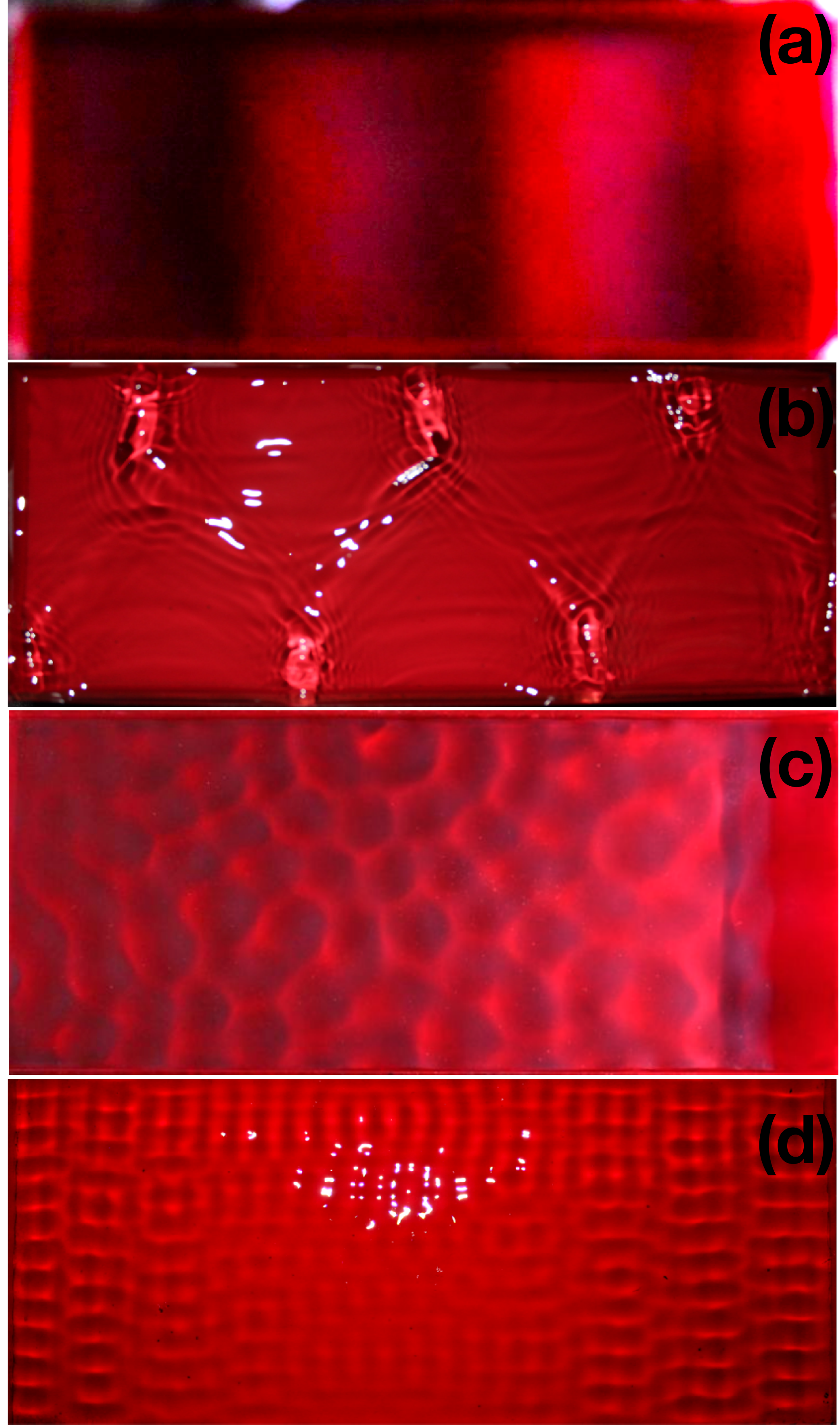}}
\caption{(a) The wave patterns for  driving frequency $\omega/\pi=7$ Hz with the sheet and (b) without the sheet. (c) The wave patterns for  driving frequency $\omega/\pi=25$ Hz with the sheet and (d) without the sheet. The size of the pictures is 33.5 cm $\times$ 14.1 cm. The wavenumber $k$ values in $m^{-1}$: (a) 61.3, (b) 79.3, (c) 216.4 and (d) 337.7. }
\label{pat}
\end{figure}

In Fig. \ref{dis}(a) we have compared the theoretical dispersion relation with the experiments. The inverse of relation (36) has been plotted as the theoretical curve for the allowed values of $k$, and has been compared to the experimental data derived from the pattern pictures. We see a good agreement. It should be noted that since we do not have the damping rates for all theoretical frequencies, we cannot calculate $\omega'$ and we present $\omega$ instead, which can cause about 5\% error in the data. We have also presented the relation for experiments in the same condition but without the covering sheet. In Fig. \ref{dis}(b) we can compare the wave numbers in two cases (with and without the sheet). We see that in low frequencies, the wave numbers are almost the same, but in high frequencies, the presence of the sheet makes $k$ values smaller i.e. larger wavelength of patterns. The comparison between the Faraday wave patterns in two cases of free surface and the covered surface with the elastic sheet can be seen in Fig. \ref{pat}. In low frequencies (a and b), the wavelength of the patterns is not very different, but in high frequencies (c and d), the difference is substantial. 

 There are some discrepancies between theory and experiment. The source of difference is not clear to us and sometimes we see one pattern for different frequencies. Maybe one reason is that the elastic sheet is not completely homogeneous and smooth. Another reason can be the waves formed on the small gap between the sheet sides and the vessel walls. It seems that all possible quantized modes cannot be produced in experiments. This has been previously reported that only a group of modes with phases in special ranges were observed in experiments \cite{DF}. Also, it was reported that the observed modes were very sensitive to the experiment details and different groups reported different modes. 

In conclusion, we studied the Faraday waves forming in a system containing a thin elastic sheet floating on water in a rectangular vessel. The threshold amplitude was calculated theoretically by adding elastic terms to the John Miles theory. The theory was compared with experimental data and a very good agreement was observed between them. The dispersion relation of the waves was obtained theoretically and compared to the experiments, showing a good agreement.

We thank L. Tuckerman for helpful discussions and comments, and S. Rasuli for helping with mior\'e technique experiments.

{}

\section{Appendix: Experimental methods}

\subsection{Driving Amplitude}

\begin{figure}[h!]
\centering
\centerline{\includegraphics[width=0.99\columnwidth]{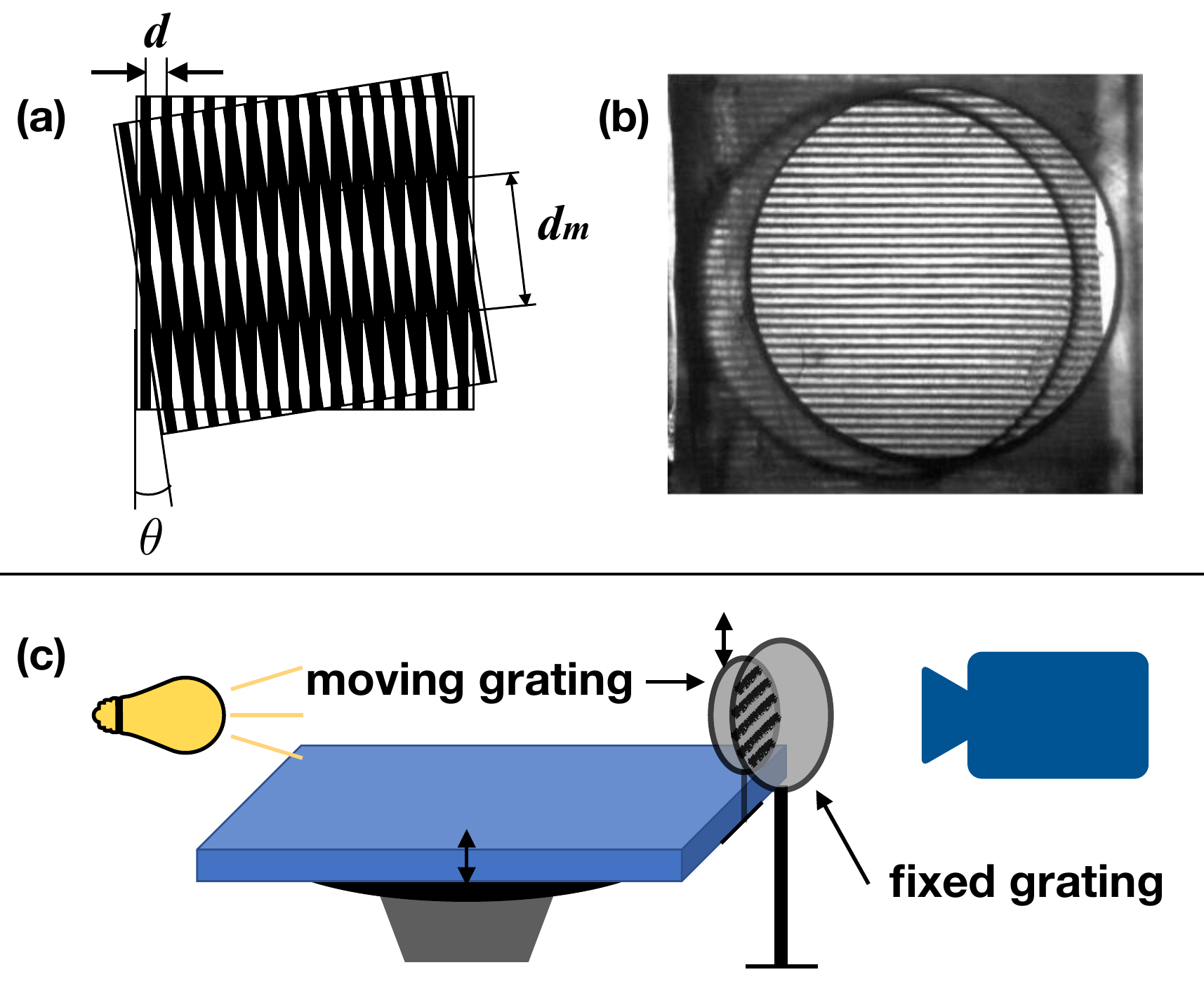}}
\caption{Moir\'e technique for measuring the oscillation amplitude.}
\label{moire1}
\end{figure}

For measuring the threshold amplitude for Faraday pattern formation, we used two optical methods. For measuring the oscillation amplitude of the vessel, we used moir\'e technique. When two gratings with the same distance $d$ between their lines are superimposed with a very small angle $\theta$, moir\'e fringes are formed with the distance $d_m$ between successive fringes. For small angle $\theta$ one obtains $d_m=d/\theta$. Thus the distance between the fringes can be very large in comparison to the distance between the grating lines. In the same way, if the gratings move a very small amount relative to each other, the displacement of the moir\'e fringes will be much larger and easier to measure, and we can obtain the respective displacement of the gratings more accurately by using the equation $\delta d = \frac{d}{d_m} \delta d_m$. This equation shows that the displacement of the lines of the gratings by the size of $\delta d$ causes the moir\'e fringes to be displaced by the size of $\delta d_m$, where $d$ is the step of gratings, and $d_m$ is the step of moir\'e fringes (Fig. \ref{moire1}). We can also use this method to measure the amplitude of vibrations of the fluid container. As we can see in Figure \ref{moire1}(b) and (c), two similar gratings with10 lines/mm were used in the experiment, one attached to the vessel and another one parallel to it outside the speaker, with a small angle between their grating lines (Fig. 1(a)). By filming the moir\'e patterns with a fast CCD camera and tracing them with a Matlab code, we were able to measure the vertical displacement of the vessel during the time with an accuracy of 50 microns, which gave us the oscillation amplitude.

\begin{figure}[h!]
\centering
\centerline{\includegraphics[width=0.99\columnwidth]{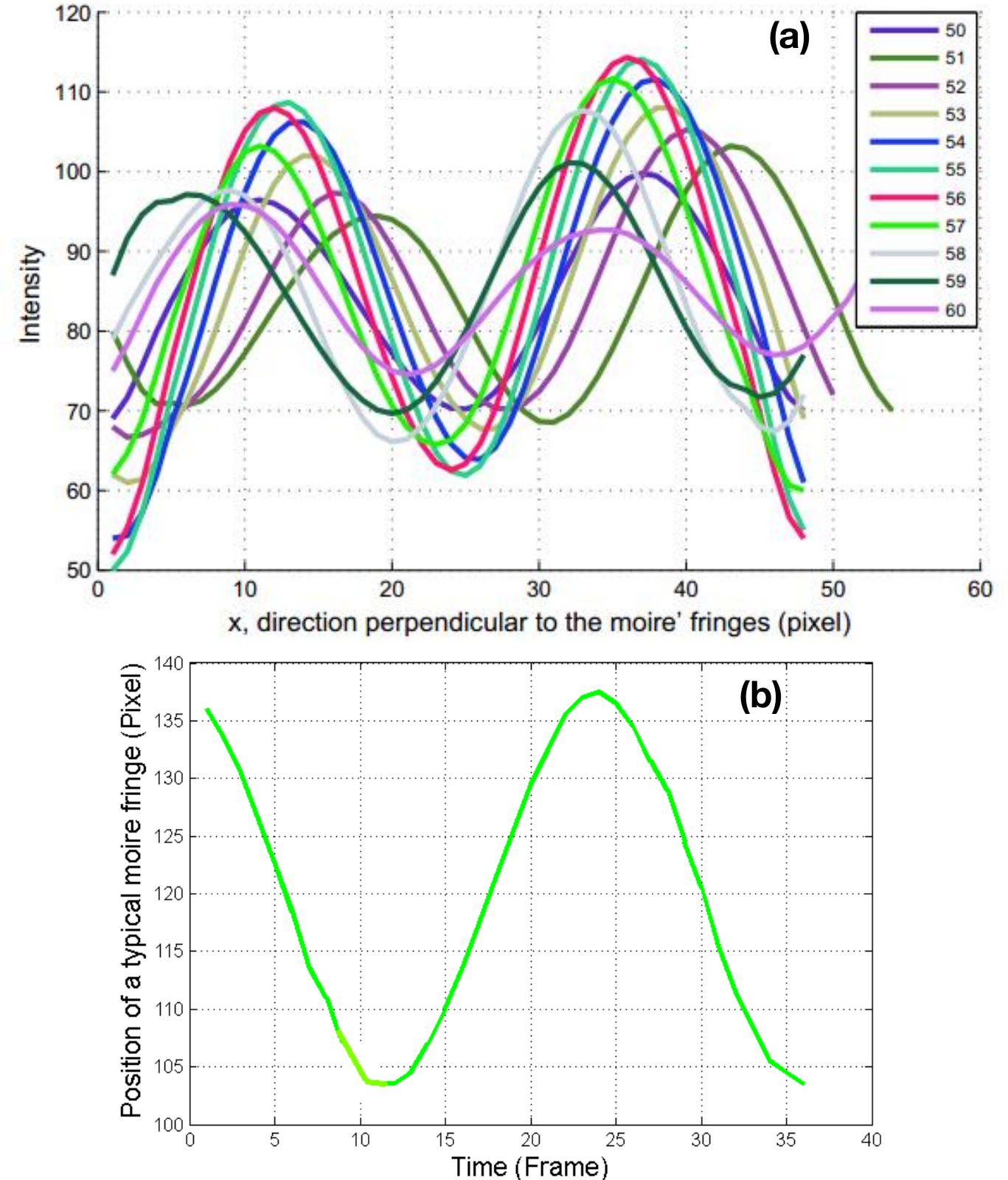}}
\caption{Moir\'{e} curves: (a) intensity of the fringe pattern vs. the horizontal coordinate $x$ in different frames, and (b) the location of one fringe tracked in successive frames. }
\label{moire2}
\end{figure}

At a constant frequency, the grating connected to the container moves parallel to the fixed grating in the vertical direction. As a result, the moir\'e fringes move in the horizontal direction. We gradually increased in the amplitude of oscillation, until the Faraday patterns were observed. At this threshold amplitude, the movement of the moir\'e fringes was recorded by a pco.1200 hs high-speed camera at a speed of 560 frames per second. Fig. \ref{moire1}(b) shows an examples of the moir\'e fringes. In order to measure the amplitude, we plotted the intensity curves of fringe pictures in different frames. The distance between two adjacent peaks (or two valleys) is the step of moir\'e fringes $d_m$ (Fig. \ref{moire2}(a)).  The displacement of a point (e.g. maximum or minimum) is followed in consecutive frames and this displacement is recorded in a graph. In fact, this diagram shows the amount of displacement of a point of the moir\'e fringes, which is caused by the movement of one of the gratings, in terms of pixels. The displacement diagram of moir\'e fringes in 35 consecutive frames at the applied frequency of 30 Hz is shown in Fig. \ref{moire2}(b). The distance between the highest and the lowest point of this graph is the maximum displacement of the moir\'e fringes.

\subsection{Damping Rate}

During the damping, the Lissajous curve became smaller. We plotted the vertical position of the laser spot as a function of time. As shown in  Fig. 3 (c, d), first we have a sinusoidal  curve (a), and after turning off the oscillations, it starts to decay exponentially. Fitting a function $y=A \sin(\omega t + \phi) \exp(-\Gamma t)$ to the experimental values, we found the damping rate $\Lambda$ for each $\omega$ in the experiments. This gave us the damping ratio $\delta=\Lambda/\omega$.

\begin{figure}[h!]
\centering
\centerline{\includegraphics[width=0.99\columnwidth]{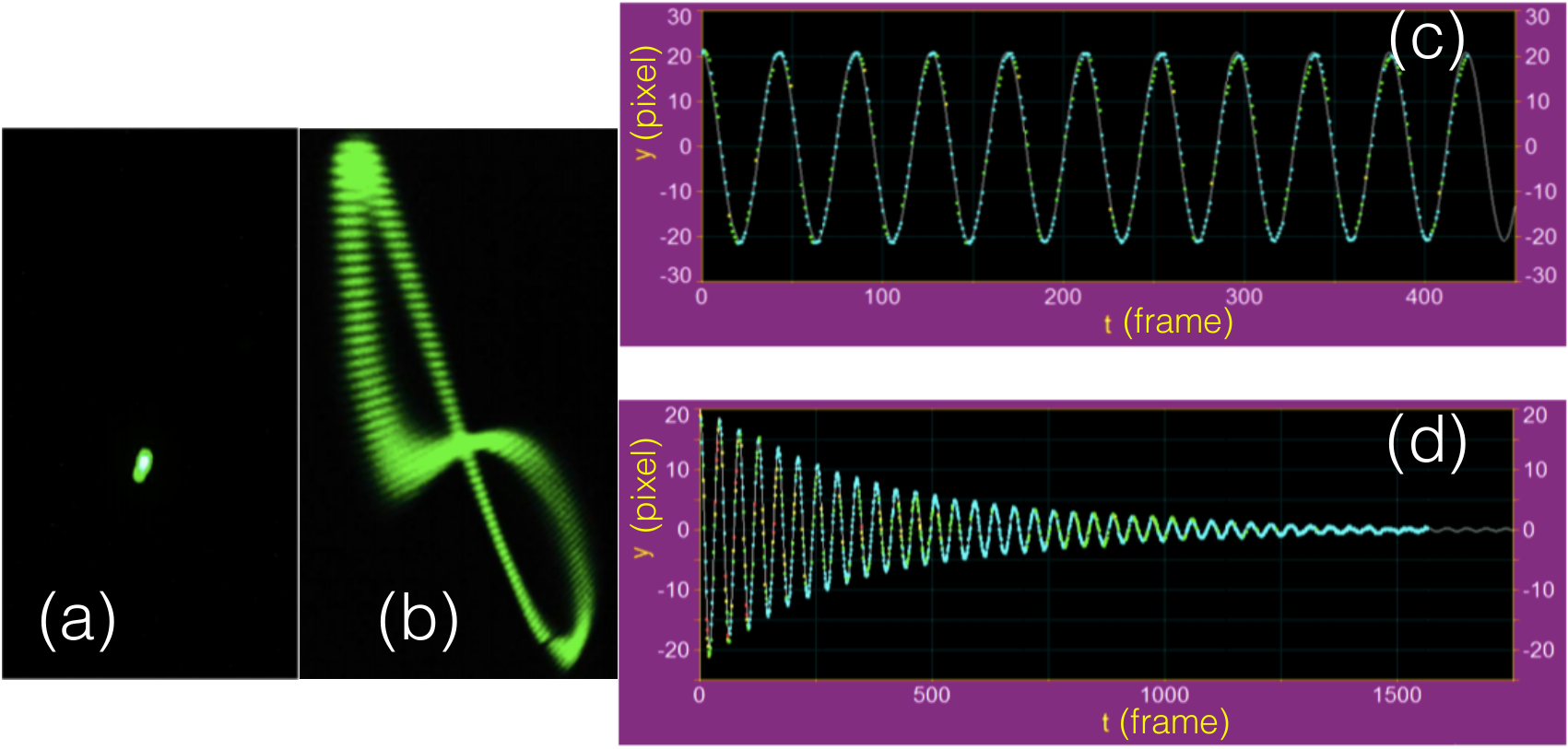}}
\caption{(a) The projection of the laser beam reflected from the surface for  an amplitude less than the threshold and no Faraday waves, and (b) an amplitude just above the threshold.(c) A typical plot of the $y$ coordinate of the laser spot on the wall as a function of time: the sinusoidal variation with time when the speaker is on; (b) the sinusoidal function times an exponentially decaying factor. Points represent the experimental values, and the curve is the fit.}
\label{Lis}
\end{figure}

\subsection{Bending Rigidity}

\begin{figure}[h!]
\centering
\centerline{\includegraphics[width=0.6\columnwidth]{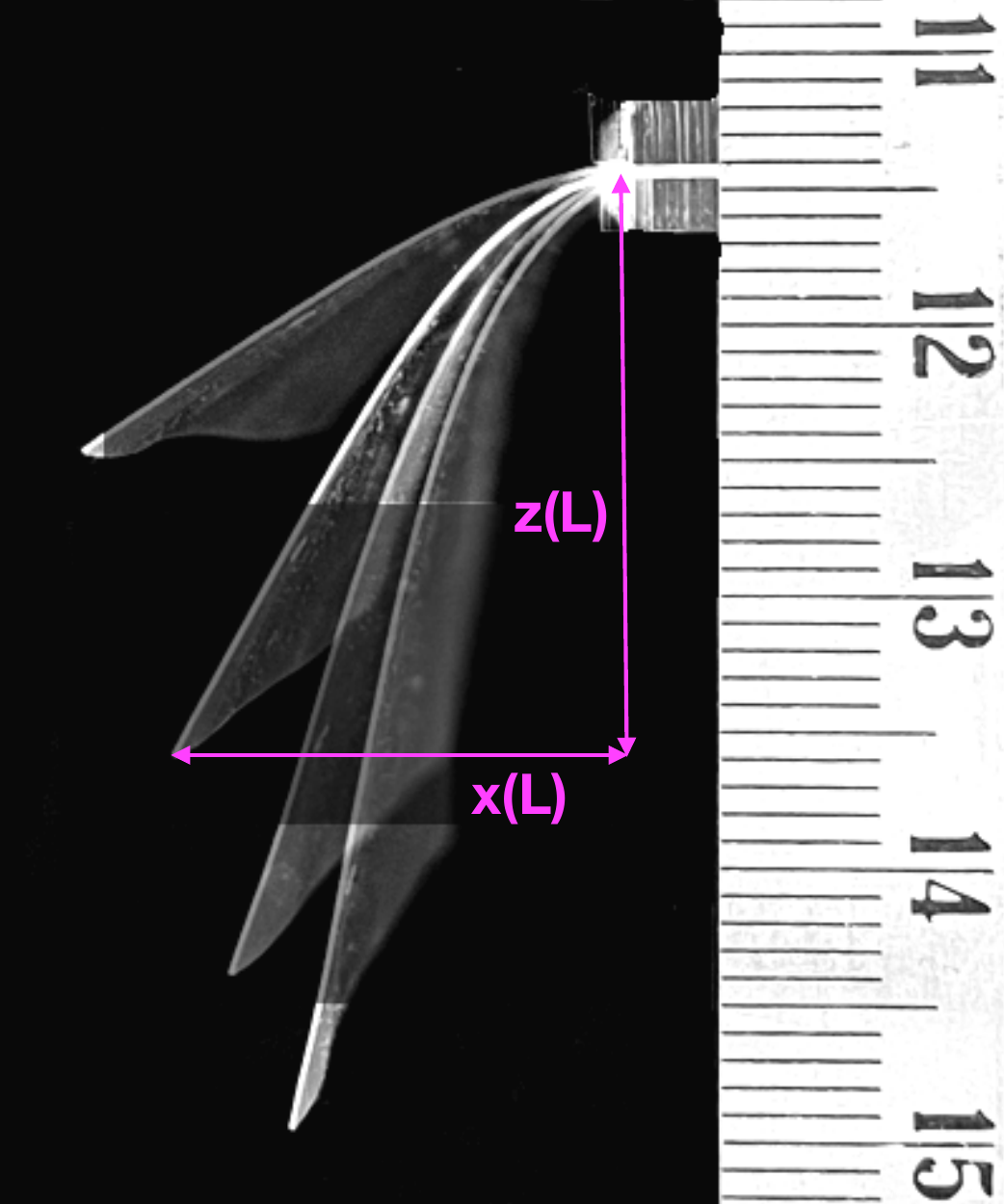}}
\caption{The method for measuring the bending rigidity of the elastic sheet. }
\label{rgdt}
\end{figure}

In order to measure the bending rigidity of the elastic sheet, we used the method introduced in reference [29] of the article. The method contains clamping different lengths $L$ of the sheet horizontally and measuring the horizontal and vertical locations of the hanging end of the sheet, $x(L)$ and $z(L)$, as shown in Fig. \ref{rgdt}. Using the relations 
\bea
\frac{L}{\xi}= a \left (\frac{z(L)}{x(L)}\right) ^{1/3}+ b  \frac{z(L)}{x(L)},  \nonumber \label{EL}\\
a=1.881, b=0.212,\\ \nonumber
\xi=\left(\frac{K}{\sigma g}\right)^{1/3},
\eea
the elastogravity length $\xi$ is calculated for each $L$ and $K$ is derived. Then an average of all obtained values of $K$ is calculated.


\begin{thebibliography}{}

\bibitem{Faraday}
M. Faraday, XVII. On a peculiar class of acoustical figures; and on certain forms assumed by groups of particles upon vibrating elastic surfaces, Philos. Trans. R. Soc. London {\bf121}, 39 (1831).

\bibitem{patterns}
K. Kumar, and K.M.S. Bajaj, Competing patterns in the Faraday experiment, Phys. Rev. E {\bf52}, R4606 (1995).

\bibitem{patterns2}
M. T. Westra, D. J. Binks, and W. Van De Water, Patterns of Faraday waves, J. Fluid Mech. {\bf496}, 1 (2003).

\bibitem{patterns3}
A. C. Skeldon, and G. Guidoboni, Pattern selection for Faraday waves in an incompressible viscous fluid, SIAM J. Appl. Math. {\bf67}, 1064 (2007).

\bibitem{patterns4}
W. S. Edwards, and S. Fauve, Patterns and quasi-patterns in the Faraday experiment, J. Fluid Mech. {\bf278}, 123 (1994).

\bibitem{patterns5}
E. Bosch, H. Lambermont, and W. van de Water, Average patterns in Faraday waves, Phys. Rev. E, {\bf49}, R3580 (1994).

\bibitem{ampl1}
P. Chen,  and J. Vinals, Amplitude equation and pattern selection in Faraday waves, Phys. Rev. E {\bf60}, 559 (1999).

\bibitem{ampl2}
D. Binks and W. van de Water, Nonlinear pattern formation of Faraday waves, Phys. Rev. Lett.  {\bf78}, 4043 (1997).

\bibitem{viscous}
J. Bechhoefer, V. Ego, S. Manneville, and B. Johnson, An experimental study of the onset of parametrically pumped surface waves in viscous fluids, J. Fluid Mech. {\bf288}, 325 (1995).   

\bibitem{viscous2}
K. Kumar, Linear theory of Faraday instability in viscous liquids, Proc. R. Soc. A: Math. Phys. Eng. Sci. {\bf452}, 1113 (1996).   

\bibitem{viscous3}
E. Cerda and E. Tirapegui, Faraday's instability for viscous fluids, Phys. Rev. Lett {\bf78}, 859 (1997).  

\bibitem{tension}
M. Perlin M and  W. W. Schultz, Capillary effects on surface waves, Annu. Rev. Fluid Mech. {\bf32}, 241 (2000).

\bibitem{tension2}
S. V. Diwakar, V Jajoo, S. Amiroudine, S. Matsumoto, R. Narayanan, F. Zoueshtiagh, Influence of capillarity and gravity on confined Faraday waves, Phys. Rev. Fluids {\bf3}, 073902 (2018). 

\bibitem{depth}
S. Ubal, M. D. Giavedoni, and F. A. Saita, A numerical analysis of the influence of the liquid depth on two-dimensional Faraday waves, Phys. Fluids {\bf15}, 3099 (2003).

\bibitem{shape}
X. Li,J. Li, X. Li, S. Liao and C. Chen, Effect of width on the properties of Faraday waves in Hele-Shaw cells, Sci. China: Phys. Mech. Astron. {\bf62}, 1 (2019).

\bibitem{shape2}
J. Yong-jun, E. Xue-Quan, and B. Wei, Nonlinear Faraday waves in a parametrically excited circular cylindrical container, Appl. Math. Mech.  {\bf24}, 1194 (2003).

\bibitem{shape3}
J. Moehlis, J. Porter and E. Knobloch, Heteroclinic dynamics in a model of Faraday waves in a square container, Physica D {\bf238}, 846 (2009).

\bibitem{shape4}
A. Wernet, C. Wagner, D. Papathanassiou, H. W. M{\"u}ller, and K. Knorr, Amplitude measurements of Faraday waves, Phys. Rev. E, {\bf63}, 036305 (2001).

\bibitem{Urcell}
T. B. Benjamin and F. J. Ursell, The stability of the plane free surface of a liquid in vertical periodic motion, Proc. R. Soc. A {\bf225}, 505 (1954).

\bibitem{Kumar}
K. Kumar and L. S. Tuckerman, Parametric instability of the interface between two fluids, J. Fluid Mech. {\bf279}, 49 (1994).

\bibitem{Miles67}
J. W. Miles, Surface-wave damping in closed basins. Proc. R. Soc. A {\bf297}, 459 (1967).

\bibitem{Miles84}
J. W. Miles, Nonlinear Faraday resonance, J. Fluid Mech. {\bf146}, 285 (1984).

\bibitem{Miles90}
D. M. Henderson and J. W. Miles, Single-mode Faraday waves in small cylinders, J. Fluid Mech. {\bf213}, 95 (1990).

\bibitem{Miles90A}
J. Miles and D. Henderson, Parametrically forced surface waves. Annual Review of Fluid Mechanics, Annu. Rev. Fluid Mech. {\bf22}, 143 (1990).

\bibitem{Miles93}
J. Miles, On faraday waves, J. Fluid Mech. {\bf248}, 671 (1993).

\bibitem{1D}
T. Khan and M. Eslamian, Experimental analysis of one-dimensional Faraday waves on a liquid layer subjected to horizontal vibrations, Phys. Fluids {\bf31}, 082106 (2019).

\bibitem{Douady}
S. Douady, Experimental study of the Faraday instability, J. Fluid Mech. {\bf221}, 383 (1990).

\bibitem{EN}
X. Jin, M. A. Xue and P. Lin, Experimental and numerical study of nonlinear modal characteristics of Faraday waves, Ocean Eng. {\bf221}, 108554 (2021).

 
\bibitem{float0}
J. W. Davys, R. J. Hosking and A. D. Sneyd, Waves due to a steadily moving source on a floating ice plate, J. Fluid Mech. {\bf158}, 269 (1985). 

\bibitem{float1}
J. Bhattacharjee and C. Guedes Soares, Flexural gravity wave over a floating ice sheet near a vertical wall, J. Eng. Math. {\bf75}, 29 (2012).

\bibitem{float2}
T. D. Williams and V. A. Squire, Scattering of flexural–gravity waves at the boundaries between three floating sheets with applications, J. Fluid Mech. {\bf569}, 113 (2006). 

\bibitem{float3}
J. C. Ono-dit-Biot, M. Trejo, E. Loukiantcheko, M. Lauch, E. Raphael, K. Dalnoki-Veress and T. Salez, Hydroelastic wake on a thin elastic sheet floating on water, Phys. Rev. Fluids {\bf4}, 014808 (2019).  
 
\bibitem{float4}
V. A. Squire, J. P. Dugan, P. Wadhams, P. J. Rottier and A. K. Liu, Annu. Of ocean waves and sea ice, Rev. Fluid Mech. {\bf27}, 115 (1995). 

\bibitem{Abou}
M. Maleki, S. M. Hashemi,  and A. Amiri, Deformation of a constrained thin elastic sheet over a flat surface due to gravity, Int. J. Non-Linear Mech. {\bf106}, 155 (2018). 

\bibitem{DF}
S. Douady and S. Fauve, Pattern selection in Faraday instability, EPL {\bf6}, 221 (1988).

\bibitem{capillary}
K. D. Nguyem Thu Lam and H. Caps, Effect of a capillary meniscus on the Faraday instability threshold, Eur. Phys. J. E {\bf34}, 1 (2011).

\end{thebibliography}
\end{document}